\documentclass[aps,superscriptaddress,amsmath,amssymb,twocolumn,showpacs,floatfix,english]{revtex4}

\usepackage{url}
\usepackage{bm}
\usepackage{graphicx}
\usepackage{amsmath}
\usepackage{amstext}
\usepackage{amssymb}
\usepackage{amsfonts}
\usepackage{amsbsy}
\usepackage{verbatim}
\usepackage{color}
\usepackage[colorlinks=true, urlcolor=blue, linkcolor=blue, citecolor=blue, pdftex]{hyperref}
\usepackage{multirow}
\usepackage{pdfpages}
\usepackage{floatrow}
\usepackage{gensymb}
\usepackage{textcomp}

\newcommand{\dagga}{{\phantom{\dagger}}}

\begin{document}

\title{Superconductivity, charge-density waves, antiferromagnetism, and phase separation in the Hubbard-Holstein model}

\author{Seher Karakuzu}
\affiliation{SISSA-International School for Advanced Studies, Via Bonomea 265, I-34136 Trieste, Italy}
\author{Luca F. Tocchio}
\affiliation{Institute for Condensed Matter Physics and Complex Systems, DISAT, Politecnico di Torino, I-10129 Torino, Italy}
\author{Sandro Sorella}
\affiliation{SISSA-International School for Advanced Studies, Via Bonomea 265, I-34136 Trieste, Italy}
\author{Federico Becca}
\affiliation{Democritos National Simulation Center, Istituto Officina dei Materiali del CNR and 
SISSA-International School for Advanced Studies, Via Bonomea 265, I-34136 Trieste, Italy}
\date{\today}

\begin{abstract}
By using variational wave functions and quantum Monte Carlo techniques, we investigate the interplay between electron-electron and 
electron-phonon interactions in the two-dimensional Hubbard-Holstein model. Here, the ground-state phase diagram is triggered by 
several energy scales, i.e., the electron hopping $t$, the on-site electron-electron interaction $U$, the phonon energy $\omega_0$, 
and the electron-phonon coupling $g$. At half filling, the ground state is an antiferromagnetic insulator for $U \gtrsim 2g^2/\omega_0$, 
while it is a charge-density-wave (or bi-polaronic) insulator for $U \lesssim 2g^2/\omega_0$. In addition to these phases, we find a 
superconducting phase that intrudes between them. For $\omega_0/t=1$, superconductivity emerges when both $U/t$ and $2g^2/t\omega_0$ 
are small; then, by increasing the value of the phonon energy $\omega_0$, it extends along the transition line between antiferromagnetic 
and charge-density-wave insulators. Away from half filling, phase separation occurs when doping the charge-density-wave insulator, while 
a uniform (superconducting) ground state is found when doping the superconducting phase. In the analysis of finite-size effects, it 
is extremely important to average over twisted boundary conditions, especially in the weak-coupling limit and in the doped case.
\end{abstract}


\maketitle

\section{Introduction}\label{sec:intro}

The challenge of understanding the interplay between electron-electron and electron-phonon interactions has stimulated an intense 
work in the condensed-matter community, since the early developments of many-body approaches to describe metals, insulators, and 
superconductors~\cite{AbrikosovBook}. Indeed, the low-temperature properties of several materials are controlled by the competition, 
or sometimes the cooperation, between these interaction terms. For example, in high-temperature superconductors, where the presence 
of a strong electron-electron correlation is irrefutable, the role of phonons could be not entirely negligible, as suggested by the 
kinks in the electron dispersion~\cite{Lanzara2001} or by the signatures of the isotope effect~\cite{Zhao1997,Gweon2004}. In alkali-metal-doped 
fullerides, a superconducting phase appears close to a Mott transition~\cite{Capone2009,Takabayashi2009}, even though they are often 
considered as phononic superconductors. This particular feature suggests that both the Coulomb repulsion and the electron-phonon coupling
are strong and cooperate to establish a strongly-correlated superconductor. Similarly, in pnictide superconductors~\cite{Cruz2008}, 
such as LaOFeAs, and in aromatic superconductors, such as potassium-intercalated picene~\cite{Mitsuhashi2010}, there are evidences that, 
apart from a moderately strong electron correlation, there is also a non-negligible coupling between electrons and lattice degrees of 
freedom. 

In a nutshell, the interplay between electron-electron repulsion and electron-phonon coupling is due to the fact that the former one 
generates spin fluctuations, which in turn mediate a non-local pairing among electrons that may give rise to $d$-wave superconductivity,
while the latter one directly mediates a local attraction among electrons, leading to an $s$-wave superconductor. In addition, a strong 
electron correlation may also lead to spin-density waves and a magnetically ordered state, which competes with superconductivity; instead,
a local attraction may also generate charge localization, i.e., charge-density-wave (CDW) or dimerized (Peierls) states. Therefore, it 
is a highly nontrivial task to obtain the properties of a system in which both interactions are relatively strong. In this respect, the 
Hubbard-Holstein model represents a prototypical example that includes these features. This model incorporates both an on-site Coulomb 
repulsion $U$ (the Hubbard term)~\cite{Hubbard1963} and a coupling $g$ between electrons and dispersionless Einstein phonons with 
energy $\omega_0$ (the Holstein terms)~\cite{Holstein1959}, as well as a kinetic term for electrons:
\begin{eqnarray}
{\cal H} = &&-t \sum_{\langle i,j \rangle, \sigma} c^\dag_{i,\sigma} c^\dagga_{j,\sigma} + {\rm H.c.} 
+U \sum_{i} n_{i,\uparrow} n_{i,\downarrow} \nonumber \\ 
&&+\omega_0 \sum_{i} b^\dag_i b^\dagga_i +g \sum_{i} n_{i} (b^\dag_i + b^\dagga_i),
\label{eq:HHmodel}
\end{eqnarray}
where $\langle i,j \rangle$ indicates nearest-neighbor sites (on a square lattice); moreover, on a given site $i$, $c^\dag_{i,\sigma}$ 
($c^\dagga_{i,\sigma}$) creates (destroys) an electron with spin $\sigma$, $b^\dag_i$ ($b^\dagga_i$) creates (destroys) a phonon, and 
$n_{i}=\sum_{\sigma} n_{i,\sigma}=\sum_{\sigma} c^\dag_{i,\sigma} c^\dagga_{i,\sigma}$ is the electron density. In analogy, we also 
define the phonon density on the site $i$ by $m_{i}=b^\dag_{i} b^\dagga_{i}$. Of course, this model gives a simplified description of 
real solids, since both the Coulomb repulsion and the electron-phonon interaction are assumed to be local. In addition, the latter term 
is modeled by coupling the lattice displacement $x_{i} \propto (b^\dag_{i}+b^\dagga_{i})$ to the electron {\it density} $n_{i}$. 
A different way to introduce the electron-phonon coupling has been considered within the Su-Schrieffer-Heeger model~\cite{Su1979}, where 
lattice displacements are coupled to the hopping term rather than to the density. While the latter case is more suited to describe 
materials with delocalized phonons, the Holstein model can be used as a good approximation for molecular solids in which there are local 
phonon modes (like, for example, fullerene doped with alkali-metal atoms).

Although the properties of the Hubbard-Holstein model depend upon $\omega_0/t$ and $g/t$ independently, it is useful to define the
quantity $\lambda=2g^2/\omega_0$, which is often considered to measure the strength of the electron-phonon coupling. This dimensionless 
coupling emerges naturally in the antiadiabatic limit where the phonons have a large energy (i.e., $\omega_0/t \to \infty$). In this 
case, the retarded interaction mediated by phonons becomes instantaneous. In fact, for $\omega_0 \gg t$ there is an exact mapping from 
the Holstein model to the negative-$U$ Hubbard model with $U_{\rm att}=-\lambda$. Therefore, the Hubbard-Holstein model reduces to the 
Hubbard model with a renormalized on-site interaction, i.e., $U_{\rm eff}=U-\lambda$. In the general case with a finite phonon energy, 
the multidimensional parameter space of the Hubbard-Holstein model (i.e., $U/t$, $g/t$, $\omega_0/t$, as well as the electron density 
$n$) leads to an extremely rich physics and various approaches have been used to understand its ground-state properties. 

In one spatial dimension, early works, based upon perturbation theory and Monte Carlo calculations, suggested that the ground state of
the Holstein model displays CDW order for any non-zero electron-phonon coupling and $\omega_0/t<\infty$~\cite{Hirsch1983,Zheng1989}. 
Instead, subsequent studies, using density-matrix renormalization group (DMRG) and Monte Carlo techniques, have highlighted the existence 
of a  gapless phase (with dominant superconducting pair correlations) for small values of the electron-phonon coupling and finite phonon 
energies, which persists also for finite values of $U/t$~\cite{Jeckelmann1998,Jeckelmann1999,Clay2005,Tezuka2007,Fehske2008,Hohenadler2013}. 
In the opposite limit  of infinite dimensions, dynamical mean-field theory (DMFT) has been employed to assess various aspects of the phase 
diagram, including the competition between superconductivity and CDW order~\cite{Freericks1993,Freericks1995,Bauer2010,Murakami2013}, 
the role of phonons in the vicinity of the Mott transition~\cite{Capone2004,Sangiovanni2005}, the verification of the Migdal-Eliashberg 
theory~\cite{Werner2007}, the polaron formation, and the existence of the isotope effect~\cite{Capone2003,Paci2005,Macridin2006}. 
The two-dimensional case has been relatively little investigated in the past. Indeed, quantum Monte Carlo techniques suffer from the 
sign problem and stable simulations can be accomplished only in few cases~\cite{Huang2003,Nowadnick2012,Johnston2013}. Therefore, the 
Hubbard-Holstein model has been mainly considered within mean-field approaches~\cite{Becca1996,Zeyer1996,Barone2008,diCiolo2009} or
by using perturbative methods~\cite{Berger1995,Hotta1997}. Variational Monte Carlo (VMC) has been also employed to assess the interplay 
between electron-electron and electron-phonon interactions in the Su-Schrieffer-Heeger model~\cite{Alder1997}. More recently, Ohgoe and 
Imada used the VMC approach to assess the ground-state phase diagram of the Hubbard-Holstein model at half filling and in its 
vicinity~\cite{Ohgoe2017}. The same variational wave functions have been implemented to study the electron-phonon coupling in multi-band 
models~\cite{Watanabe2015}.

One important aspect in the phase diagram of the Hubbard-Holstein model is the nature of the transition between CDW (bi-polaronic) and 
Mott insulators at half filling and the possibility that a metallic/superconducting phase may intrude in between~\cite{Takada1996,Takada2003}. 
In the antiadiabatic limit $\omega_0/t \to \infty$, given the mapping from the Holstein model to the negative-$U$ Hubbard model, one should 
expect a direct transition between an antiferromagnetic (Mott) insulator, that is stable for $U>\lambda$, and a CDW insulator, that is stable 
for $U<\lambda$. However, in this limit, the CDW state is degenerate with an $s$-wave superconductor, because of the SU(2) pseudo-spin 
symmetry of the negative-$U$ Hubbard model (this fact leads to peculiar ground-state properties, with both broken translational symmetries, 
i.e., CDW order, and gapless excitations). By contrast, for any finite values of the phonon energy $\omega_0$, there is no reason for 
having a direct transition between the two insulating states and an intermediate phase may emerge. In one dimension, the existence of a 
metallic phase, with strong superconducting correlations, has been reported by DMRG studies~\cite{Clay2005,Tezuka2007,Fehske2008}, with a 
clear evidence that the intermediate region broadens with increasing the phonon energy (up to $\omega_0/t \approx 5$). Instead, DMFT 
calculations showed contradictory results, with either a direct transition between CDW and Mott insulators~\cite{Bauer2010} or the presence 
of a small intermediate phase~\cite{Murakami2013}. Also in two dimensions the situation is not conclusive, since only few calculations have 
been afforded~\cite{Nowadnick2012,Johnston2013}, where some evidence for the emergence of an intermediate metallic phase has been suggested 
at finite temperatures. In addition, away from half filling, the sign problem is so strong that it prevents one from performing any stable 
simulation. Therefore, alternative approaches are highly desirable. One possibility is to define suitable wave functions that can be treated 
within the VMC technique. In this spirit, Ohgoe and Imada have recently extended the ``many-variable'' VMC method to include phonon degrees 
of freedom~\cite{Ohgoe2017,Ohgoe2014}, showing evidence in favor of a metallic (with weak superconducting correlations) phase between the 
CDW and Mott insulators at half filling. In addition, they highlighted the presence of phase separation when doping the CDW insulator. 
Instead, the ground state is uniform when doping the metallic phase.

In this paper, we present further VMC calculations that are based upon different wave functions and smart average over twisted boundary
conditions (denoted by TABC) in order to reduce size effects. Indeed, when imposing periodic or antiperiodic boundary conditions, there are 
very large size effects, especially for small values of $\omega_0/t$ and in the doped case, preventing us from reaching definitive conclusions 
in the thermodynamic limit. Thanks to TABC, we give a clear evidence that a superconducting phase is present between the CDW and the 
antiferromagnetic insulators and that its stability region broadens when increasing the phonon energy. Finally, phase separation is found 
when doping the CDW state, while a uniform (superconducting) phase is observed by doping the uniform ground state. 

The paper is organized as follows: in Sec.~\ref{sec:methods}, we show the variational wave function and briefly discuss the Monte Carlo
methods that have been used; in Sec.~\ref{sec:results}, we present the numerical results; finally, in Sec.~\ref{sec:conclusions}, we 
draw our conclusions.

\section{Wave functions and methods}\label{sec:methods}

In this section, we first describe the variational wave function that has been used in the numerical calculations. Then, we briefly 
discuss the updating scheme that has been implemented within the VMC method for the phononic degrees of freedom (for the electrons, we
use standard updating schemes~\cite{BeccaBook}). Finally, we show the TABC procedure to reduce size effects and we highlight 
advantages and disadvantages of the VMC method.

The wave function is given by the so-called Jastrow-Slater state that is defined by:
\begin{equation}\label{eq:wf}
|\Psi\rangle = {\cal J}_{ee} {\cal J}_{pp} {\cal J}_{ep} {\cal P}_{N_e} |\Phi_{e}\rangle \otimes |\Phi_{p}\rangle.
\end{equation}
Here, $|\Phi_{e}\rangle$ is the ground state of an auxiliary (quadratic) Hamiltonian that contains electron hopping and (singlet) 
pairing:
\begin{eqnarray}
&& {\cal H}_{\rm aux} = -t \sum_{\langle i,j \rangle,\sigma} c^\dag_{i,\sigma} c^\dagga_{j,\sigma} + {\rm H.c.}
- \mu \sum_{i,\sigma} c^\dag_{i,\sigma} c^\dagga_{i,\sigma} \nonumber \\
&& + \sum_{i,\sigma} (-1)^{X_i+Y_i} \left [ \Delta_{\rm CDW} + \Delta_{\rm AF} (-1)^{\bar{\sigma}} \right ] 
c^\dag_{i,\sigma} c^\dagga_{i,\sigma} \nonumber \\
&& +\Delta_{\rm SC} \sum_{i} c^\dag_{i,\uparrow} c^\dag_{i,\downarrow} + {\rm H.c.},
\label{eq:auxham}
\end{eqnarray}
where $\mu$, $\Delta_{\rm CDW}$, $\Delta_{\rm AF}$, and $\Delta_{\rm SC}$ are parameters that are optimized in order to minimize the 
variational energy~\cite{Sorella2005}, ${\bf R}_i=(X_i,Y_i)$ indicates the coordinates of the site $i$ in the square lattice, and 
$\bar{\sigma}=+1\, (-1)$ for up (down) electrons. The auxiliary Hamiltonian of Eq.~(\ref{eq:auxham}) is rather flexible to describe 
states with (i) CDW order (when $\Delta_{\rm CDW} \ne 0$ and $\Delta_{\rm AF}=\Delta_{\rm SC}=0$), (ii) antiferromagnetic N\'eel order 
(when $\Delta_{\rm AF} \ne 0$ and $\Delta_{\rm CDW}=\Delta_{\rm SC}=0$), and (iii) superconducting order (when $\Delta_{\rm SC} \ne 0$ 
and $\Delta_{\rm CDW}=\Delta_{\rm AF}=0$). Moreover, states with coexisting orders are also possible. This Hamiltonian can be easily 
diagonalized to define the uncorrelated electronic state $|\Phi_{e}\rangle$, which has the following form:
\begin{equation}\label{eq:wfelec}
|\Phi_{e}\rangle = \exp \left ( \sum_{i,j} f_{i,j} c^\dag_{i,\uparrow} c^\dag_{j,\downarrow} \right )|0\rangle,
\end{equation}
where the pairing function $f_{i,j}$ depends upon the variational parameters of the auxiliary Hamiltonian. The total number of 
electrons is fixed to $N_e$ by the projector ${\cal P}_{N_e}$. 

The uncorrelated phononic part is then given by:
\begin{equation}
|\Phi_{p}\rangle = \sum_{N_b} \frac{\left ( e^{\zeta} \; b^\dag_{k=0} \right )^{N_b}}{{N_b}!} |0\rangle, 
\end{equation}
where $b^\dag_{k=0}=1/\sqrt{N}\sum_{i} b^\dag_{i}$ creates a phonon in the $k=0$ momentum state ($N$ is the number of sites). Here, 
$N_b$ denotes the total number of phonons. Since the number of phonons is not conserved by the Hubbard-Holstein Hamiltonian of 
Eq.~(\ref{eq:HHmodel}), $|\Phi_{p}\rangle$ has components on subspaces with any value of $N_b$; then, $\zeta$ is a variational parameter 
that plays the role of a fugacity. Denoting by $|m_{1},\dots,m_{N}\rangle$ the (normalized) configuration with $m_{i}$ phonons on the 
site $i$, the uncorrelated phononic wave function can be written as:
\begin{equation}\label{eq:wfphon}
|\Phi_{p}\rangle = \sum_{m_1,\dots,m_N} \frac{e^{\zeta \sum_{i} m_{i}}}{\sqrt{m_1! \dots m_N!}} |m_{1},\dots,m_{N}\rangle.
\end{equation}

Finally, ${\cal J}_{ee}$, ${\cal J}_{pp}$, and ${\cal J}_{ep}$ are density-density Jastrow factors for the electron-electron,
phonon-phonon, and electron-phonon correlations, respectively:
\begin{eqnarray}
&& {\cal J}_{ee} = \exp \left ( -\frac{1}{2} \sum_{i,j} v^{ee}_{i,j} n_{i} n_{j} \right ), \label{eq:Jee} \\
&& {\cal J}_{pp} = \exp \left ( -\frac{1}{2} \sum_{i,j} v^{pp}_{i,j} m_{i} m_{j} \right ), \label{eq:Jpp} \\
&& {\cal J}_{ep} = \exp \left ( - \sum_{i,j} v^{ep}_{i,j} n_{i} m_{j} \right ), \label{eq:Jep}
\end{eqnarray}
where $v^{ee}_{i,j}$, $v^{pp}_{i,j}$, and $v^{ep}_{i,j}$ are pseudo-potentials, including the on site terms, that are taken to be 
translationally invariant, i.e., they depend only upon the Euclidean distance $|{\bf R}_{i}-{\bf R}_{j}|$. They can be optimized (each one 
independently) to reach the optimal variational {\it Ansatz}~\cite{Sorella2005}. Here, all the pseudo-potentials are taken to be symmetric 
in the exchange $i \leftrightarrow j$. By a full optimization of the Jastrow factors, we find that the phonon-phonon correlations only 
give a marginal improvement in the energy and, therefore, they are not employed. By contrast, the electron-phonon term is fundamental to 
obtain an accurate description when $g/t$ and $\omega_0$ are finite. As for the Hubbard model, the electron-electron Jastrow factor is 
crucial to reproduce the correct low-energy behavior of the ground state~\cite{Capello2005,Capello2006}.

%%%%%%%%%%%%%%%%%%%%%%%%%%%%%%%%%%%%%%%%%%%%%
\begin{figure}
\includegraphics[width=1.0\columnwidth]{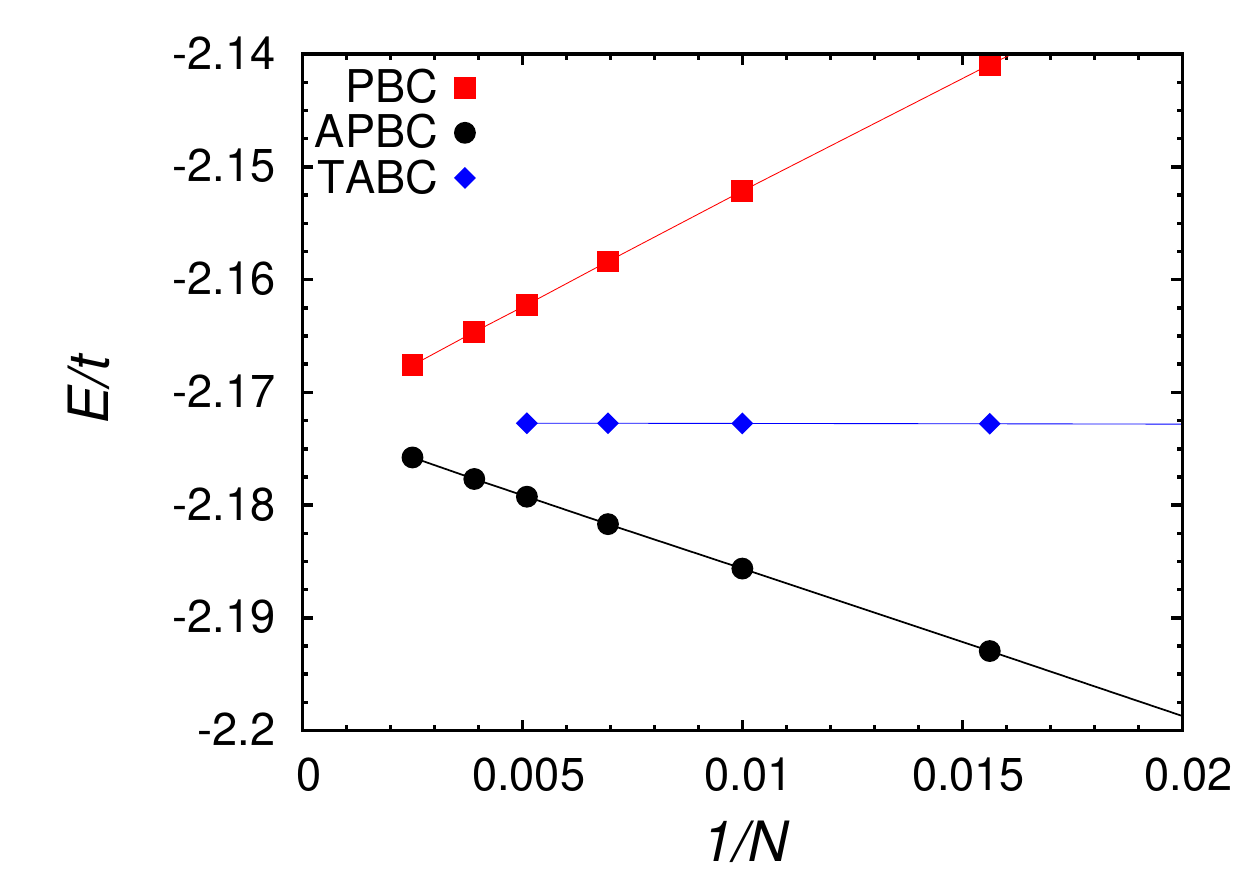}
\caption{\label{fig:boundary1}
Size scaling of the energy per site at half filling (with $\omega_0/t=1$, $\lambda/t=0.98$, and $U=0$) for periodic-periodic (red squares), 
periodic-antiperiodic (black circles), and twisted average (blue diamonds) boundary conditions. Errorbars are smaller than the size of the 
symbols.}
\end{figure}

\begin{figure}
\includegraphics[width=1.0\columnwidth]{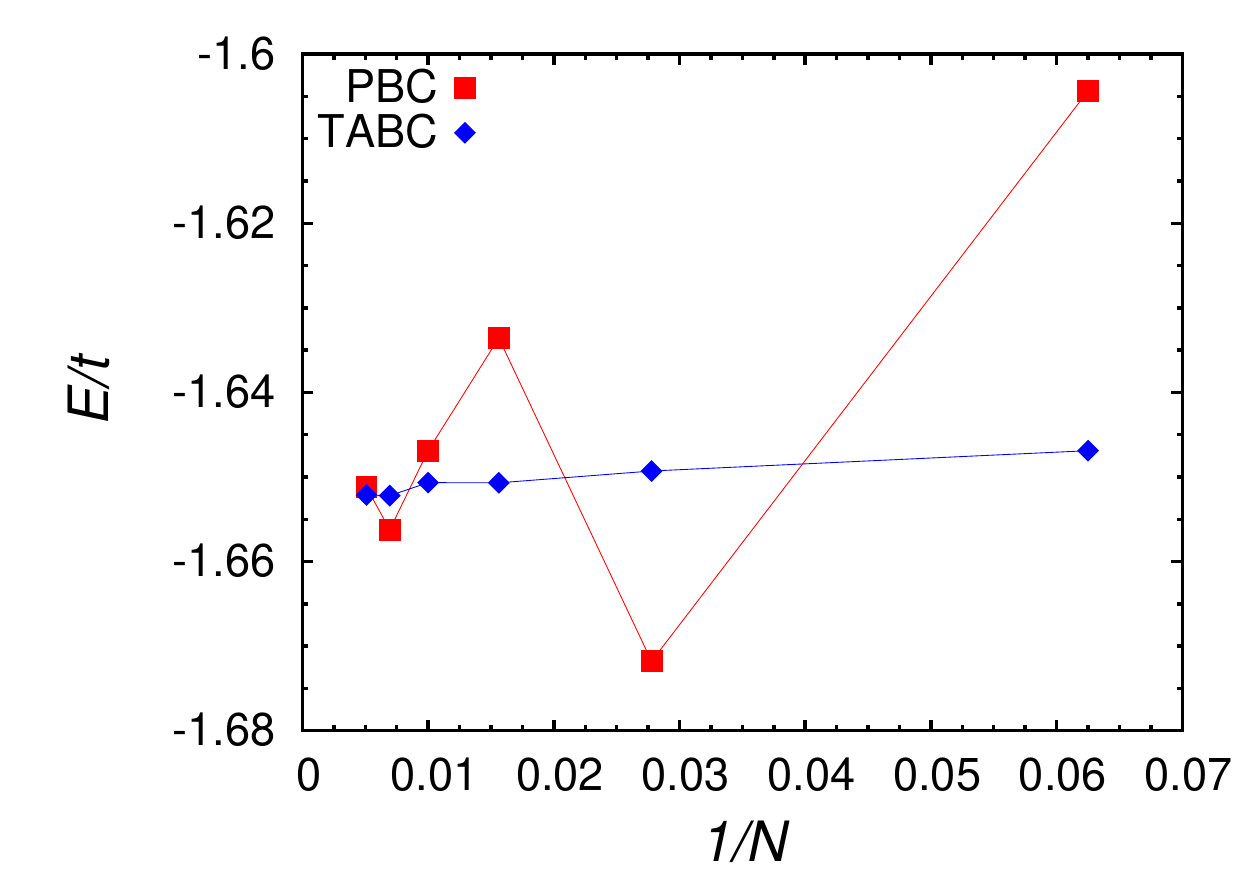}
\caption{\label{fig:boundary2}
Size scaling of the energy per site at quarter filling (with $\omega_0/t=1$, $\lambda/t=2$, and $U=0$) for periodic-periodic (red squares) 
and twisted average (blue diamonds) boundary conditions. Errorbars are smaller than the size of the symbols.}
\end{figure}
%%%%%%%%%%%%%%%%%%%%%%%%%%%%%%%%%%%%%%%%%%%%%

The configuration space that is sampled along the Markov chain is defined by specifying both electron and phonon occupations on each
site (i.e., we work in a basis in which the number of phonons $m_{i}$ is specified on each lattice site, as well as the number of 
up- and down-spin electrons $n_{i,\sigma}$). In our case, where the uncorrelated phononic part is given by Eq.~(\ref{eq:wfphon}) and
contains a single variational parameter $\zeta$, we do not need to include any cutoff in the number of phonons. By contrast, Ohgoe and 
Imada~\cite{Ohgoe2017,Ohgoe2014} used a more involved parametrization of the phonon wave function, with several parameters (i.e., one
for each boson number); therefore, they considered a cutoff in the maximum number of phonons. Moreover, along the Markov chain, they 
sampled the displacement $x_i$, using a different electron-phonon Jastrow factor, which couples the electron density $n_{j}$ to $x_{i}$ 
(rather than to the phonon number $m_{i}$). Finally, also the electronic part used in Refs.~\cite{Ohgoe2017,Ohgoe2014} is slightly 
different from the one that is employed here: they do not obtain $|\Phi_{e}\rangle$ from the auxiliary Hamiltonian~(\ref{eq:auxham}), 
but perform a full optimization of the pairing function $f_{i,j}$ of Eq.~(\ref{eq:wfelec}). In our opinion, this procedure may be 
problematic, especially for large sizes, since one must deal with several variational parameters (i.e., $O(N)$ for a translationally 
invariant case) and the optimization of the long-range tail of the pairing function can be difficult within a stochastic approach. 
An advantage of our parametrization is that the nature of the wave function is transparent from the optimized values of the parameters, 
e.g., obtaining a finite $\Delta_{\rm CDW}$ immediately implies that the state displays CDW order. We also mention that the present 
approach allows us to easily detect metastable phases, with given physical properties: for example, by fixing $\Delta_{\rm AF}=0$, 
we are able to obtain the best paramagnetic state, even in a region where the ground state is antiferromagnetically ordered. 

When using the wave function of Eq.~(\ref{eq:wf}), the Metropolis algorithm can be easily implemented to propose a change in the phononic 
configuration. Indeed, let us consider the case in which one phonon is created/destroyed at site $l$, i.e., $m_{i} \to m_{i} \pm \delta_{il}$.
Then, in order to compute the Metropolis acceptance probability, it is necessary to evaluate the following ratio:
\begin{equation}
{\cal R}^{\pm}_l = \frac{\langle m_1,\dots,m_l \pm 1,\dots,m_N|{\cal J}_{pp} {\cal J}_{ep}|\Psi_{p}\rangle}
{\langle m_1,\dots,m_l,\dots,m_N|{\cal J}_{pp} {\cal J}_{ep}|\Psi_{p}\rangle},
\end{equation}
which can be explicitly given by the expressions of the uncorrelated phononic state~(\ref{eq:wfphon}) and the Jastrow 
factors~(\ref{eq:Jpp}) and~(\ref{eq:Jep}):
\begin{eqnarray}
{\cal R}^{+}_l = \frac{e^{\zeta}}{\sqrt{m_l+1}} \exp \left [ -\sum_{i} ( v^{ep}_{i,l} n_{i}+v^{pp}_{i,l} m_{i} )-v^{pp}_{l,l} \right ], \\
{\cal R}^{-}_l = \sqrt{m_l} \; e^{-\zeta} \exp \left [ +\sum_{i} ( v^{ep}_{i,l} n_{i}+v^{pp}_{i,l} m_{i} )-v^{pp}_{l,l} \right ].
\end{eqnarray}

Finally, let us discuss the TABC method to reduce the size effects. Our calculations are performed on square clusters with $N=L \times L$
sites. In most cases, periodic (or anti-periodic) boundary conditions are employed on both the Hamiltonian and the variational wave 
function (i.e., the auxiliary Hamiltonian defined above). However, strong size effects may be present, due to a large correlation length.
It has been proposed that a selected twist in the boundary condition, or an average over different boundary conditions, may improve the 
convergence to the thermodynamic limit~\cite{Lin2001,Dagrada2016}. On the lattice, by explicitly indicating the coordinates of the site 
in the creation operators (i.e., $c^\dag_{i,\sigma} \to c^\dag_{{\bf R}_i,\sigma}$, where ${\bf R}_i=(X_i,Y_i)$ denotes the coordinates 
of the site $i$ in the lattice), twisted boundary conditions correspond to impose:
\begin{eqnarray}
&& c^\dag_{{\bf R}_i+{\bf L_x},\sigma} = e^{i\theta^{\sigma}_{x}} c^\dag_{{\bf R}_i,\sigma}, \label{eq:twistx} \\
&& c^\dag_{{\bf R}_i+{\bf L_y},\sigma} = e^{i\theta^{\sigma}_{y}} c^\dag_{{\bf R}_i,\sigma}, \label{eq:twisty}
\end{eqnarray}
where ${\bf L_x}=(L,0)$ and ${\bf L_y}=(0,L)$ are the vectors that define the periodicity of the cluster; $\theta^{\sigma}_{x}$ and 
$\theta^{\sigma}_{y}$ are two phases in $[0,2\pi)$ that determine the twists along $x$ and $y$ directions. In order to preserve 
time-reversal invariance, we must impose that $\theta^{\uparrow}_{x}=-\theta^{\downarrow}_{x}$, and similarly for the $y$ term. Then,
for each choice of $\theta \equiv (\theta^{\uparrow}_{x},\theta^{\downarrow}_{x},\theta^{\uparrow}_{y},\theta^{\downarrow}_{y})$, we 
define the many-body wave function $|\Psi_{\theta}\rangle$, which is obtained from the auxiliary Hamiltonian~(\ref{eq:auxham}) with
twisted boundary conditions (notice that the Jastrow factors are not affected by the twist, since they contain density-density 
correlations).

%%%%%%%%%%%%%%%%%%%%%%%%%%%%%%%%%%%%%%%%%%%%%
\begin{figure}
\includegraphics[width=0.98\columnwidth]{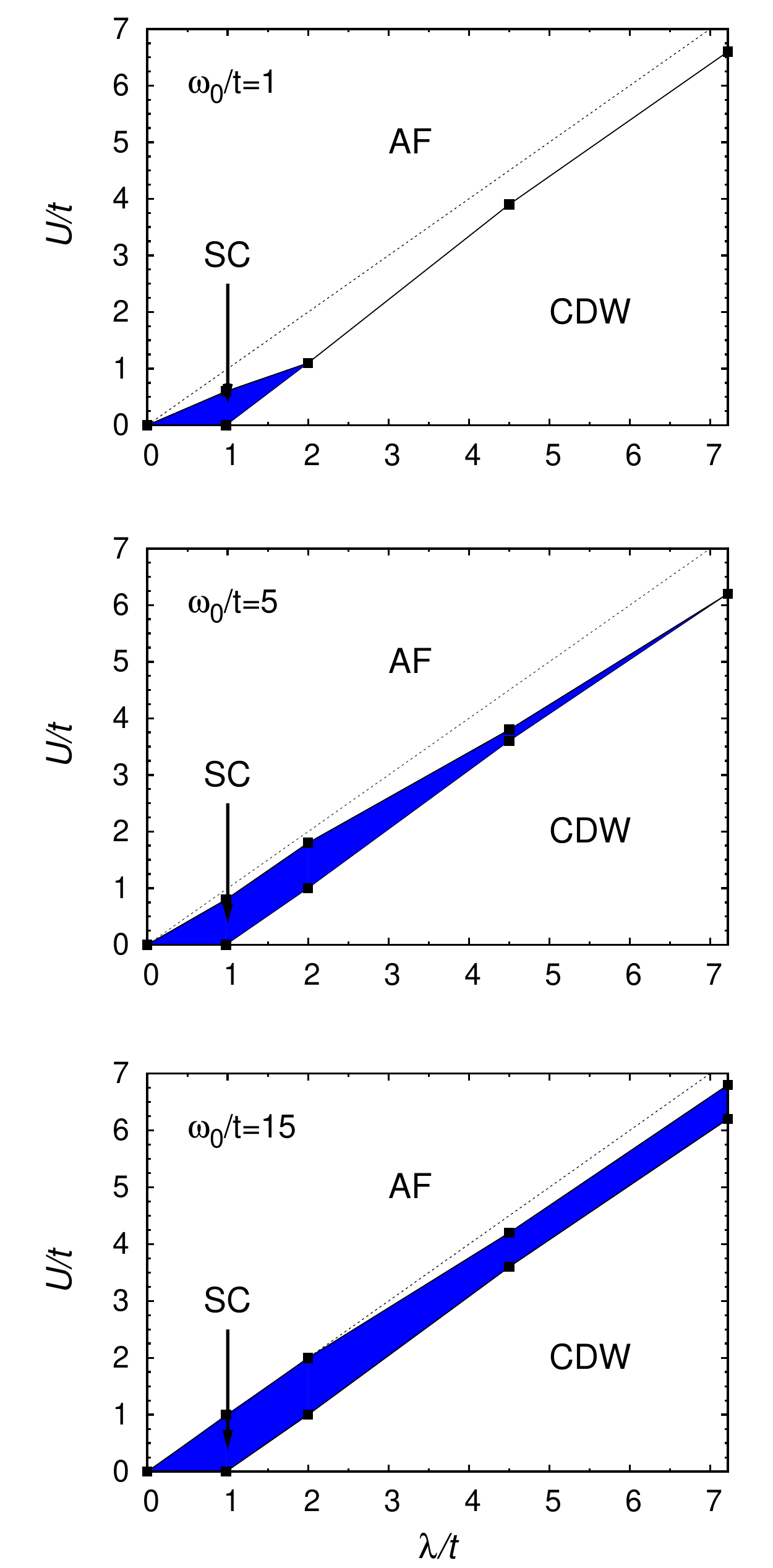}
\caption{\label{fig:phased}
Ground-state phase diagram in the $(\lambda/t,U/t)$ plane for the Hubbard-Holstein model at $\omega_0/t=1$ (upper panel), $\omega_0/t=5$ 
(middle panel), and $\omega_0/t=15$ (lower panel). Antiferromagnetic (AF), charge-density-wave (CDW), and superconducting (SC) phases 
are present, the latter one being marked in blue. The calculations have been performed on the $12 \times 12$ cluster with TABC. 
The dotted line $U=\lambda$ is also marked for reference.}
\end{figure}

\begin{figure}
\includegraphics[width=1.0\columnwidth]{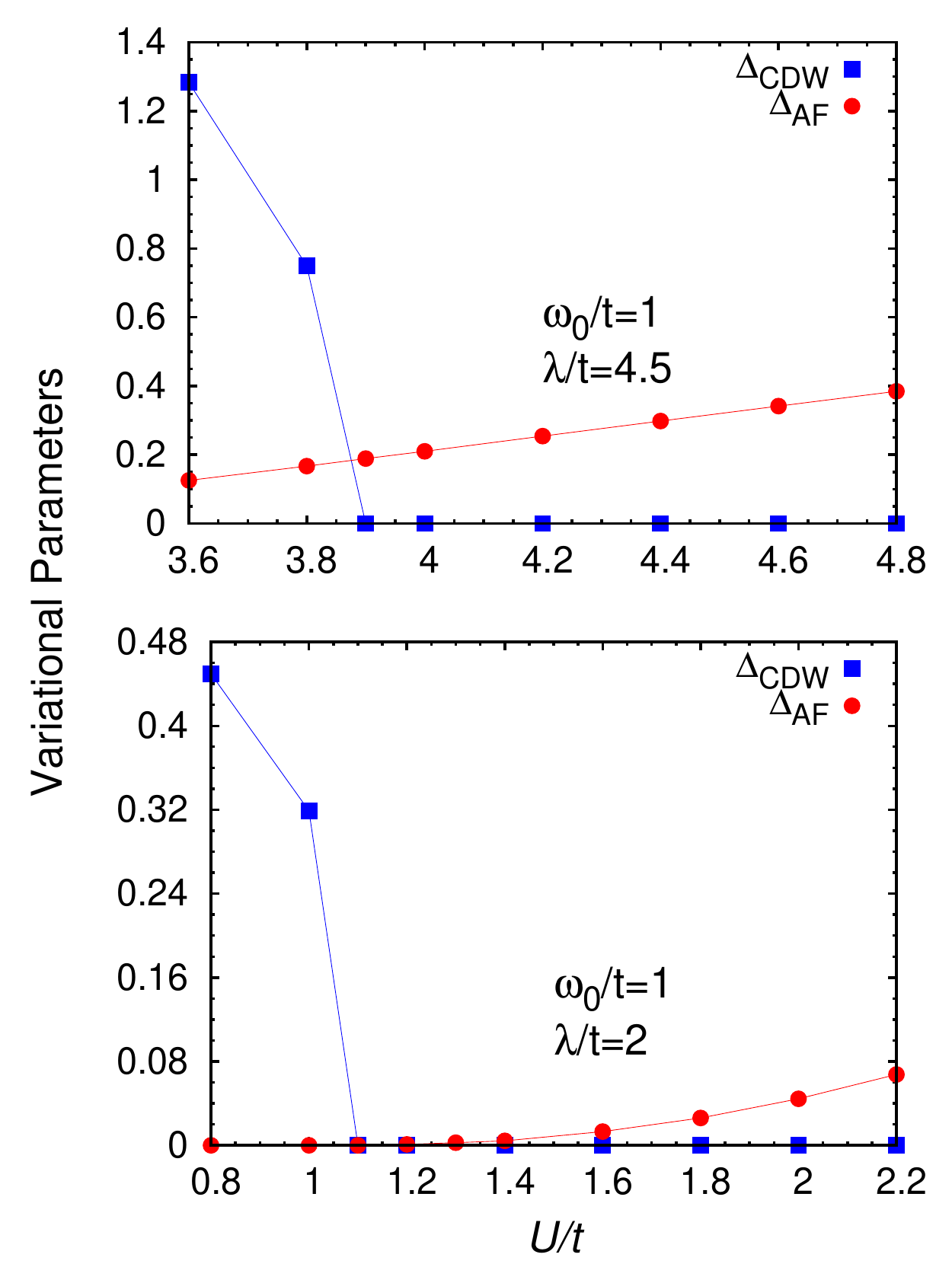}
\caption{\label{fig:paramw1}
Antiferromagnetic (red circles) and charge-density-wave (blue squares) parameters for the case with $\omega_0/t=1$. The cases with 
$\lambda/t=4.5$ (upper panel), where a first-order phase transition between these two insulators is present, and with $\lambda/t=2$ (lower 
panel), where a continuous phase transition takes place, are reported. The calculations have been performed on the $12 \times 12$ cluster 
with TABC and error bars are smaller than the size of the symbols.}
\end{figure}
%%%%%%%%%%%%%%%%%%%%%%%%%%%%%%%%%%%%%%%%%%%%%%%%

In TABC, an average over a large number $N_{\theta}$ of phases (typically $N_{\theta}=576$ points in the Brillouin zone) is considered 
in order to evaluate the expectation value of the Hamiltonian or any other operator ${\cal O}_{\theta}$, which in general depends upon 
the twist through Eq.~(\ref{eq:twistx}) and~(\ref{eq:twisty}):
\begin{equation}\label{eq:kaverage}
\langle {\cal O} \rangle = \frac{1}{N_{\theta}} \sum_{\theta} \frac{\langle \Psi_{\theta}| {\cal O}_{\theta} |\Psi_{\theta}\rangle}
{\langle \Psi_{\theta}|\Psi_{\theta}\rangle}.
\end{equation}
By imposing that all the twists share the same variational parameters, we can reach much faster the thermodynamic limit. For example,
by considering uncorrelated (i.e., mean-field) wave functions, we verified that this procedure allows us to get the thermodynamic results 
even when using a small cluster. Moreover, within a Monte Carlo optimization, the average of Eq.~(\ref{eq:kaverage}) is very conveniently 
implemented, since the statistical error decreases with $1/\sqrt{N_{\theta}}$ and, therefore, several boundary conditions can be considered 
without any extra computational cost. In Fig.~\ref{fig:boundary1}, we show the size scaling of the energy per site when applying the TABC 
procedure at half filling (for $U=0$, $\lambda/t=0.98$, and $\omega_0/t=1$), in comparison with the standard cases with periodic-periodic 
and periodic-antiperiodic boundary conditions. In all three cases, the optimized variational wave functions have $\Delta_{\rm SC} \ne 0$ 
(and $\Delta_{\rm CDW}=\Delta_{\rm AF}=0$) and the extrapolated values are all consistent (within few error-bars), giving $E/t=-2.1725(1)$. 
Away from half filling, size effects become even more pronounced and TABC are crucial to extract an accurate value in the thermodynamic limit. 
In Fig.~\ref{fig:boundary2}, we report the case at quarter filling (with $U=0$, $\lambda/t=2$, and $\omega_0/t=1$). Here, periodic-periodic 
boundary conditions give scattered results, while averaging over twisted boundary conditions gives rise to a rather smooth extrapolation to 
$E/t=-1.652(1)$. The important message is that, while the cases with fixed boundary conditions possess huge size effects and require large 
clusters to reach accurate results in the thermodynamic limit, a remarkable flat size scaling is obtained by using TABC, thus allowing us to 
consider relatively small clusters in our numerical simulation, with small finite-size errors.

%%%%%%%%%%%%%%%%%%%%%%%%%%%%%%%%%%%%%%%%%%%%%%%%5
\begin{figure}
\includegraphics[width=1.0\columnwidth]{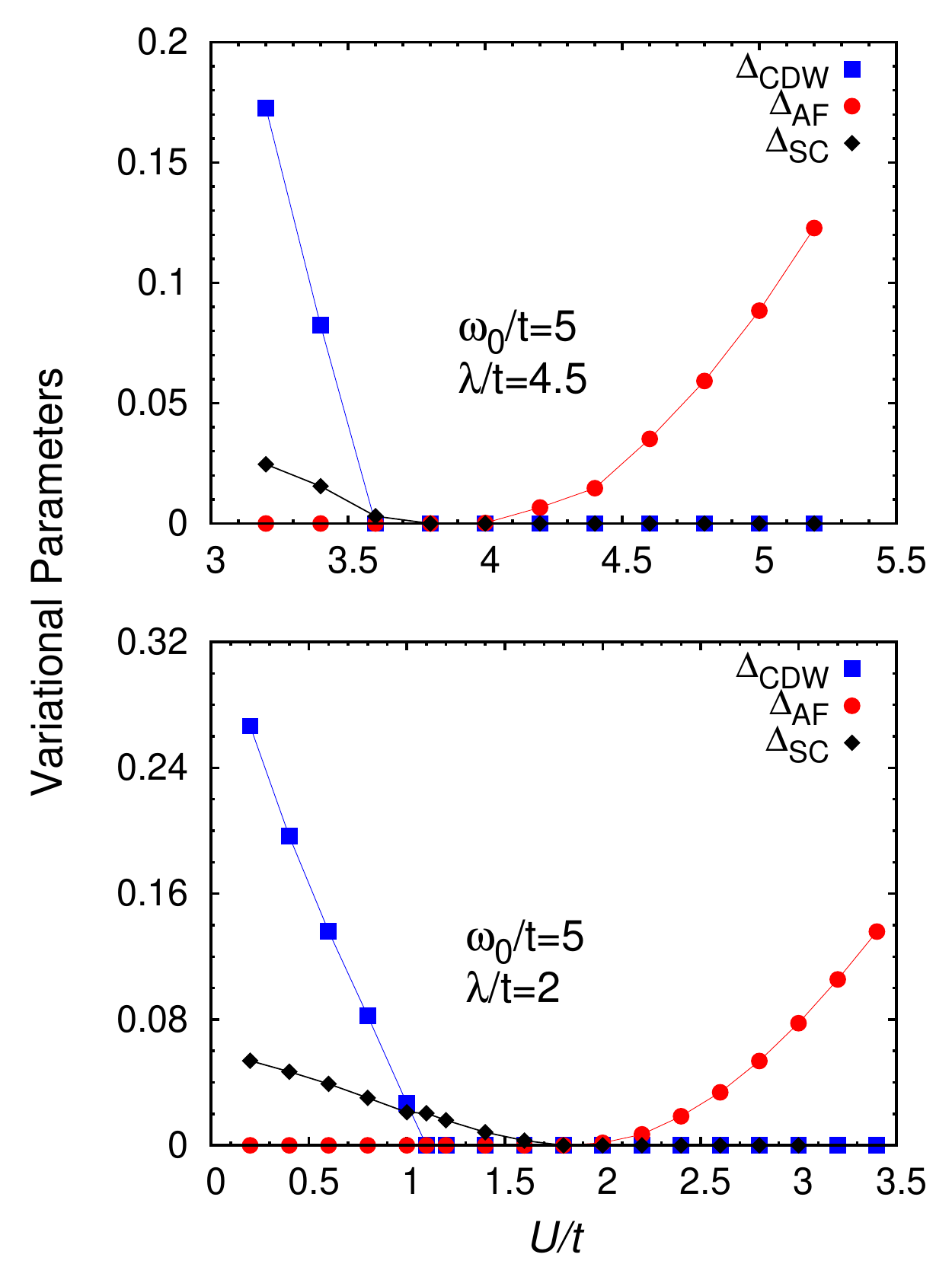}
\caption{\label{fig:paramw5}
Antiferromagnetic (red circles), charge-density-wave (blue squares), and superconducting (black diamonds) parameters for the case with 
$\omega_0/t=5$. The cases with $\lambda/t=4.5$ (upper panel) and $\lambda/t=2$ (lower panel) are reported. In both cases, there is a small 
region where the ground state is superconducting with no charge-density-wave order. The calculations have been performed on the $12 \times 12$ 
cluster with TABC and error bars are smaller than the size of the symbols.}
\end{figure}

\begin{figure}
\includegraphics[width=1.0\columnwidth]{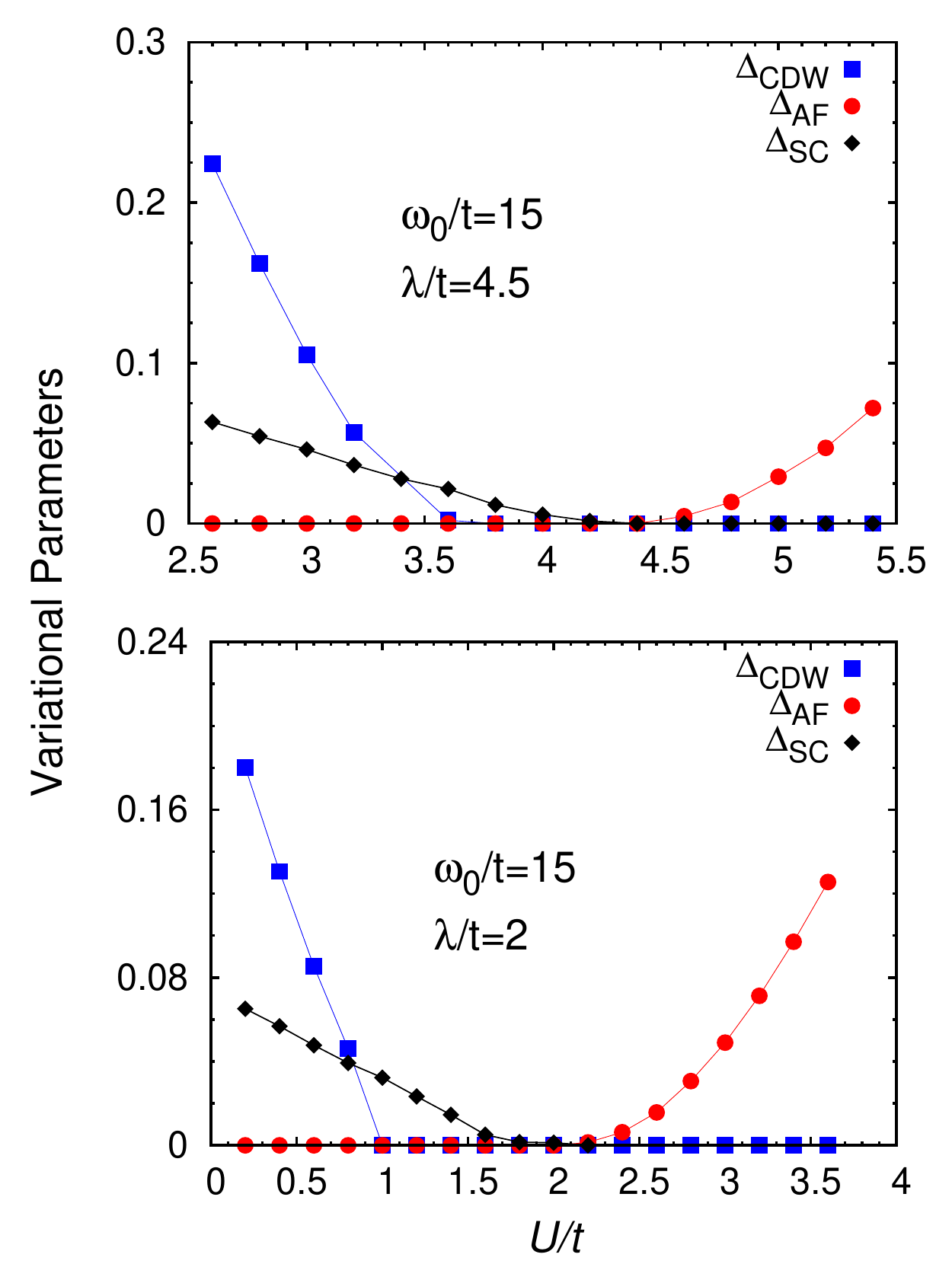}
\caption{\label{fig:paramw15}
Antiferromagnetic (red circles), charge-density-wave (blue squares), and superconducting (black diamonds) parameters for the case with 
$\omega_0/t=15$. The cases with $\lambda/t=4.5$ (upper panel) and $\lambda/t=2$ (lower panel) are reported. In this cases, there is a 
substantial region where the ground state is superconducting with no charge-density-wave order. The calculations have been performed on 
the $12 \times 12$ cluster with TABC and error bars are smaller than the size of the symbols.}
\end{figure}
%%%%%%%%%%%%%%%%%%%%%%%%%%%%%%%%%%%%%%%%%%%%%

We would also like to briefly discuss the advantages and disadvantages of computing ground-state properties by means of a Monte Carlo sampling 
over variational wave functions. The main advantage is that strongly-correlated states may be treated beyond perturbative approaches. For 
example, the physical properties (e.g., energy and correlations functions) of the simple Gutzwiller wave function can be assessed without 
considering the Gutzwiller approximation~\cite{Brinkman1970,Vollhardt1984}. In order to compute expectation values over variational states, 
a Monte Carlo sampling is necessary, thus leading to statistical errors, which, however, can be safely kept under control (i.e., they scale to 
zero by increasing the length of the simulation). The energy computed with variational Monte Carlo gives an upper bound to the exact value, 
thus providing a criterion to judge the quality of the variational states. Moreover, it is possible to assess quite large clusters, with all 
relevant spatial symmetries (translations, rotations, and reflections) preserved. The main disadvantage is that it is difficult to quantify 
the systematic errors, which are introduced by the choice of the trial state.

\section{Results}\label{sec:results}

In this section, we start by showing our numerical results for the half-filled case $n=N_e/N=1$ and then move to the doped region with 
$n<1$. 

\subsection{Half-filled case}

In Fig.~\ref{fig:phased}, we show the ground-state phase diagram for three values of $\omega_0/t$, i.e., $\omega_0/t=1$, $5$, and $15$, 
at half filling. Here, we identify three different phases. For large electron-electron interaction, the lowest-energy state has 
long-range antiferromagnetic order (namely the uncorrelated part of the electronic wave function has $\Delta_{\rm AF} \ne 0$). The 
stability region of this phase is approximately bounded (from below) by the line $U=\lambda$, for all the values of the phonon 
energies. This is a remarkable feature, which has been already obtained by different approaches, especially in one dimension by 
DMRG~\cite{Clay2005,Tezuka2007,Fehske2008} and in two dimensions by VMC~\cite{Ohgoe2017}. In fact, this is expected in the antiadiabatic
limit where $\omega_0/t \to \infty$, but there are no simple reasons that it should also hold for finite (and relatively small) values
of $\omega_0$. For large electron-phonon coupling, the ground state is a CDW insulator, where doubly-occupied sites (doublons) and empty 
sites (holons) form a checkerboard pattern; this charge modulation is accompanied by a considerable phonon ``dressing'', namely a large 
number of phonons are present on top of doublons, while no phonons are present on empty sites. This phonon cloud gives a drastic reduction 
of the kinetic energy of electrons, which hardly hop around in the lattice. Finally, there is an intermediate superconducting phase (with 
pairing correlations that increase by increasing $\omega_0/t$) that intrudes between the previous insulators. For small values of the 
phonon energy, it is limited to a narrow region for small couplings, while for intermediate values of $\omega_0/t$, it expands inside the 
region where $U<\lambda$. By further increasing $\omega_0/t$, the superconducting correlations get stronger and stronger, eventually 
pervading the whole CDW region. It should be mentioned that, when $\omega_0/t \to \infty$, the CDW state returns into the game, being 
degenerate with the superconducting one (due to the emerging SU(2) pseudo-spin symmetry that connects the superconducting and the CDW 
states). In practice, we cannot recover an exact degeneracy between these two states, since the density-density Jastrow factor favors the 
superconducting one for very large phonon energies.

For $\omega_0/t=1$ and large $\lambda/t$, the transition between the CDW and the antiferromagnetic insulators is first order, since both 
wave functions can be stabilized also when the competitor gives the lowest variational energy. For example, the variational parameters 
$\Delta_{\rm AF}$ and $\Delta_{\rm CDW}$ across the transition for $\lambda/t=4.5$ are reported in Fig.~\ref{fig:paramw1}. By decreasing 
$\lambda/t$, the local minima disappear and the transition appears to be continuous. For $\lambda/t=2$, which approximately corresponds to 
the tip of the superconducting region, CDW and antiferromagnetic parameters vanish for $U/t \approx 1.2$, see Fig.~\ref{fig:paramw1}. For 
smaller values of $\lambda/t$, a superconducting phase can be stabilized for small enough electron-electron repulsions, with a small but
clearly finite pairing term $\Delta_{\rm SC}$. We would like to remark that for $U=0$, within our variational approach (implemented with 
TABC), we obtain a finite value of the electron-phonon coupling $\lambda/t \approx 1$, separating superconducting and CDW phases. In the
non-interacting limit, at the density where the Van Hove singularity occurs (i.e., at half filling, when only the nearest-neighbor hopping $t$ is present), both the 
particle-hole and the particle-particle susceptibilities diverge as $\ln^{2}(t/\Lambda)$~\cite{Hlubina1997}, where $\Lambda$ is an infrared 
cutoff. The former one has a larger prefactor with respect to the superconducting one, thus implying that, within the mean-field approach, 
an infinitesimal interaction will lead to CDW. However,  bare susceptibilities may lead to an incorrect prediction and it is 
important to go beyond this approximation. Our variational calculations should capture the correct qualitative picture, i.e., the presence 
of an extended superconducting region below a given $\lambda/t$, even though we cannot exclude subtle finite-size effects that could be 
particularly difficult to control even by using TABC. In this respect, our results contrast with recent calculations obtained by using a
finite-temperature quantum Monte Carlo method~\cite{Weber2017}.

When increasing the phonon energy, the region of stability for the superconducting phase broadens, intruding between the two insulators 
also when $\lambda/t$ is large. In Figs.~\ref{fig:paramw5} and~\ref{fig:paramw15}, we report the behavior of the variational parameters 
$\Delta_{\rm AF}$, $\Delta_{\rm CDW}$, and $\Delta_{\rm SC}$ for $\omega_0/t=5$ and $15$. In the intermediate region, both $\Delta_{\rm AF}$ 
and $\Delta_{\rm CDW}$ are vanishing, while $\Delta_{\rm SC}$ is finite. Notice that $\Delta_{\rm SC}$ is also finite inside the insulating 
CDW region. This fact does not lead to a super-solid ground state (i.e., a superconducting state with CDW order), since the presence of a
finite $\Delta_{\rm CDW}$ is associated with a gap in the excitation spectrum (we determine whether the system is metallic or insulating  
by looking at the density-density correlations, see for instance Ref.~\onlinecite{Tocchio2011}). By contrast, this result may be ascribed to 
the fact that superconducting and CDW solutions become degenerate for $\omega_0/t \to \infty$, and, therefore, at the variational level, 
some energy gain can be obtained by mixing superconductivity and CDW order, even when the phonon energy is large but finite. Finally, we 
remark that the transition between the antiferromagnetic insulator and the superconductor appears to be continuous, i.e., both $\Delta_{\rm AF}$ 
and $\Delta_{\rm SC}$ vanish (approximately) at the same point. This is a particularly remarkable and unexpected feature, since $s$-wave 
superconductivity and local moments are not compatible.

\subsection{Doped case}

Let us now move to the doped case, for which we want to assess the stability toward phase separation. When doping an antiferromagnet, 
phase separation could appear for small hole concentrations, as found in the repulsive-$U$ Hubbard model, whenever the loss in the 
magnetic contribution to the total energy is larger than the gain due to the kinetic part. The presence of phase separation in the 
repulsive-$U$ Hubbard model has been confirmed by different methods, even if its extension as a function of $U$ is still 
controversial~\cite{Macridin2006b,Aichhorn2007,Chang2008,Misawa2014,Sorella2015,Tocchio2016,Simkovic2017}. In order to highlight the 
possible presence of phase separation in the Hubbard-Holstein model, it is very useful to consider the so-called energy per 
hole~\cite{Emery1990}:
\begin{equation}\label{eq:emery}
\epsilon(\delta) = \frac{E(\delta)-E(0)}{\delta},
\end{equation}
where $E(\delta)$ is the energy per site at hole doping $\delta=1-n$. For a uniform phase, $\epsilon(\delta)$ has a monotonically 
increasing behavior with increasing $\delta$ from $0$ to $1$; by contrast, phase separation is marked by the presence of a minimum of
$\epsilon(\delta)$ on any finite-size clusters and a flat behavior (up to $\delta_c$) in the thermodynamic limit. These facts can be
easily understood by considering that $\epsilon(\delta)$ represents the slope of the line joining $(0,E(0))$ to $(\delta,E(\delta))$ and 
that, in a stable uniform phase $E(\delta)$ is a convex function, while phase separation implies (after Maxwell construction) a linear
behavior of $E(\delta)$ up to $\delta_c$.

%%%%%%%%%%%%%%%%%%%%%%%%%%%%%%%%%%%%%%%%%%%%%
\begin{figure}
\includegraphics[width=1.0\columnwidth]{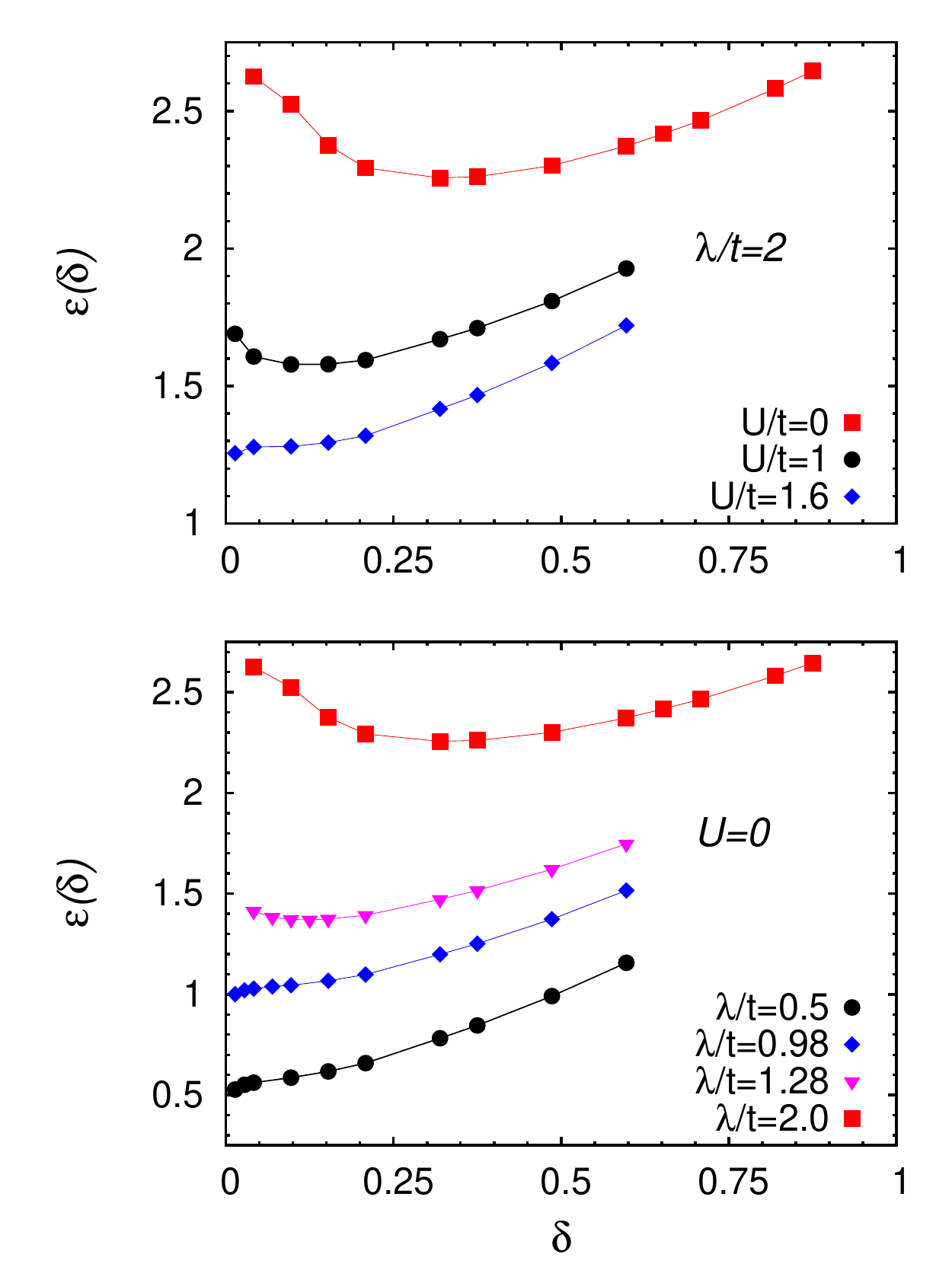}
\caption{\label{fig:phasesep1}
Energy per hole of Eq.~(\ref{eq:emery}) for $\lambda/t=2$ and various values of $U/t$ (upper panel) and for $U=0$ and various values of 
$\lambda/t$ (lower panel). In both cases $\omega_0/t=1$ and calculations are performed on the $12 \times 12$ cluster with TABC. 
Errorbars are smaller than the size of the symbols.}
\end{figure}

\begin{figure}
\includegraphics[width=1.0\columnwidth]{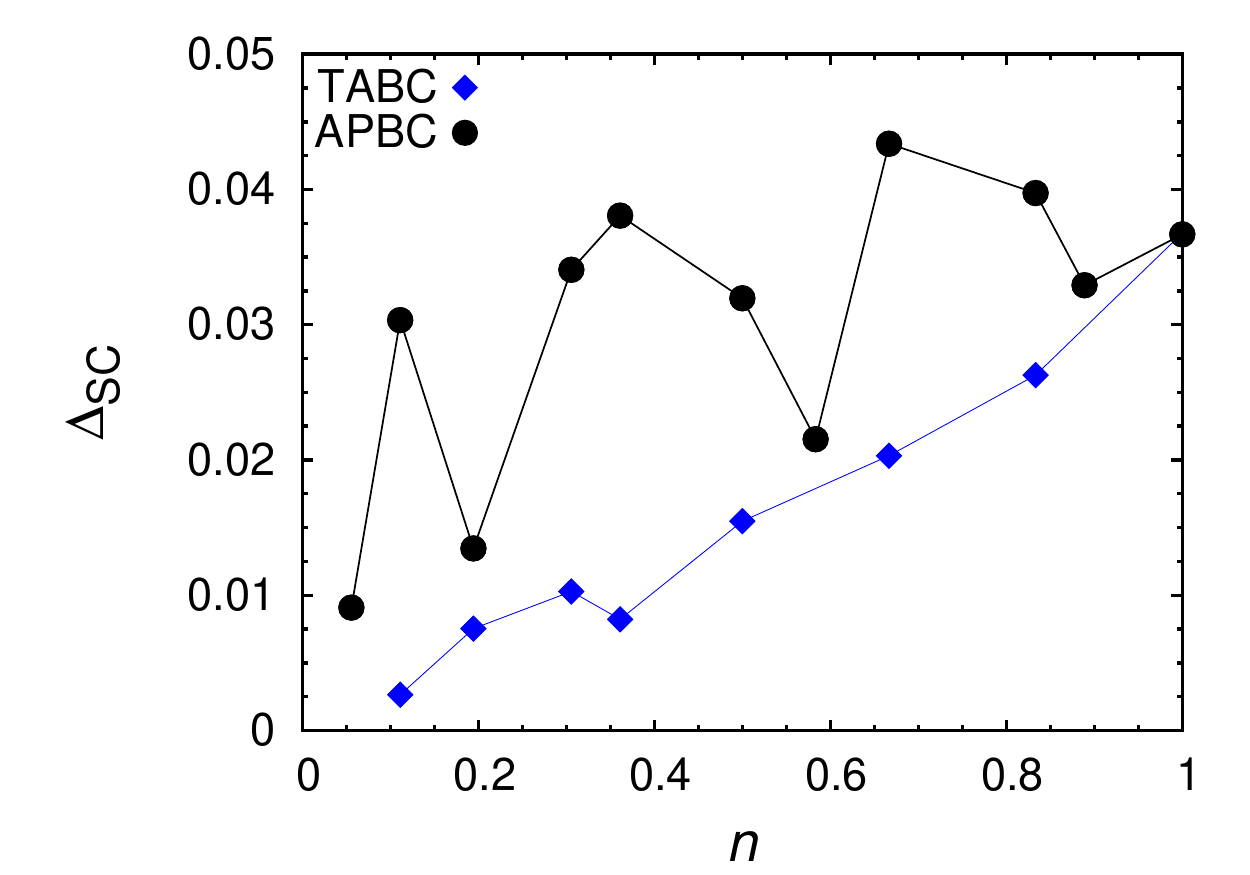}
\caption{\label{fig:deltavsn}
Variational parameter $\Delta_{\rm SC}$ as a function of the electron density $n$, computed with periodic-antiperiodic (black circles), and 
twisted average (blue diamonds) boundary conditions on the $12 \times 12$ cluster. Here, $\omega_0/t=15$ and $\lambda/t=0.98$. Errorbars are 
smaller than the size of the symbols.}
\end{figure}
%%%%%%%%%%%%%%%%%%%%%%%%%%%%%%%%%%%%%%%%%%%%%

The results of the energy per hole are shown in Fig.~\ref{fig:phasesep1}. First of all, we discuss the case with $U=0$, $\omega_0/t=1$, and 
finite $\lambda$ (lower panel). Here, the system does not phase separate for small values of the electron-phonon coupling, i.e., when doping 
the superconducting phase at half filling. Most importantly, the ground state remains superconducting also when the electron density is
$n<1$. This fact is most evident when $\omega_0/t$ is large enough, since the superconducting signal is rather small in the adiabatic 
limit and increases with $\omega_0/t$. In order to show this feature, we present the results for $\omega_0/t=15$ in Fig.~\ref{fig:deltavsn}. 
As for the half-filled case, in order to get smooth results when the electron density is varied, it is fundamental to consider TABC, since
fixed boundary conditions (here, periodic-antiperiodic ones) give rise to a strongly scattered behavior. By contrast, when entering into 
the CDW phase for large values of $\lambda/t$, a small hole doping leads to a charge instability, with the region where phase separation is 
obtained increasing with $\lambda$. We remark that, within TABC, the results show a smooth behavior that is not obtained by using fixed boundary 
conditions. As for the case of an antiferromagnetic phase, also in the presence of CDW order the injection of few mobile holes that damage 
the charge periodicity is not compensated by a kinetic energy gain. Thus, phase separation appears for sufficiently small hole doping. 
In Fig.~\ref{fig:phasesep1}, we also show the results for $\lambda/t=2$ and various values of $U/t$ (upper panel). Here, the electron-electron 
repulsion opposes to the electron-phonon coupling, leading to a reduction of phase separation until it eventually disappears above a 
critical value of $U/t$ (by further increasing the electron-electron repulsion, antiferromagnetism settles down at half filling, thus 
leading again to phase separation, as discussed in the positive-$U$ Hubbard model).

\begin{figure}
\includegraphics[width=1.0\columnwidth]{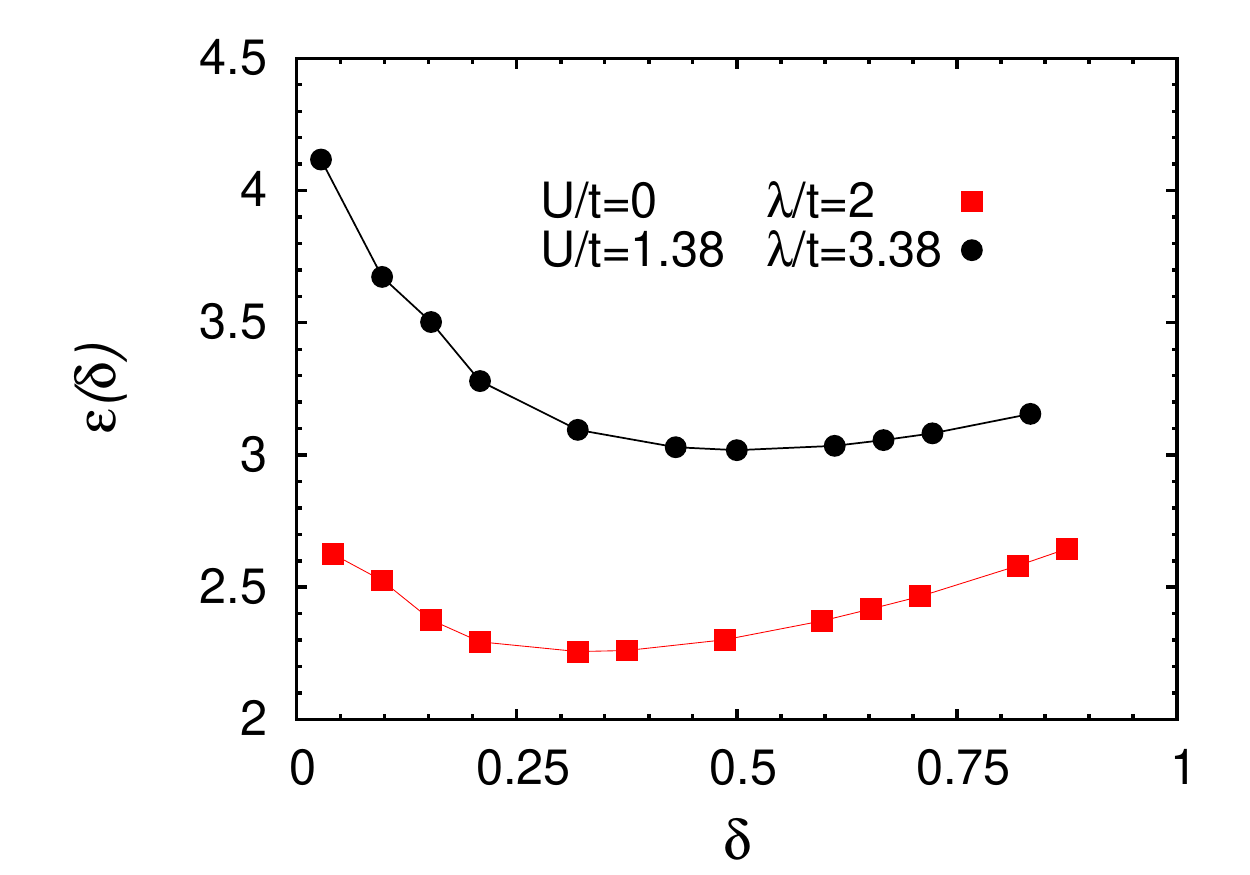}
\caption{\label{fig:phasesep2}
Energy per hole of Eq.~(\ref{eq:emery}) for $U=0$ and $\lambda/t=2$ (red squares), compared with $U/t=1.38$ and $\lambda/t=3.38$ (black 
circles), that corresponds to the same value of the effective interaction $U_{\rm eff}=U-\lambda$. In both cases $\omega_0/t=1$ and calculations
are performed on the $12 \times 12$ cluster with TABC. Errorbars are smaller than the size of the symbols.}
\end{figure}

In Fig.~\ref{fig:phasesep2}, we further show that, at finite values of the phonon energy, the extent of phase separation depends upon the 
actual values of both $U$ and $\lambda$. Indeed, we observe that, for $\omega_0/t=1$, phase separation is more pronounced for $U/t=1.38$ 
and $\lambda/t=3.38$ than for $U=0$ and $\lambda/t=2$, even if both cases would give the same effective interaction $U_{\rm eff}=U-\lambda$. 
This fact can be explained by the presence, at half filling, of a larger CDW parameter in the former case with respect to the latter one. 

\begin{figure}
\includegraphics[width=1.0\columnwidth]{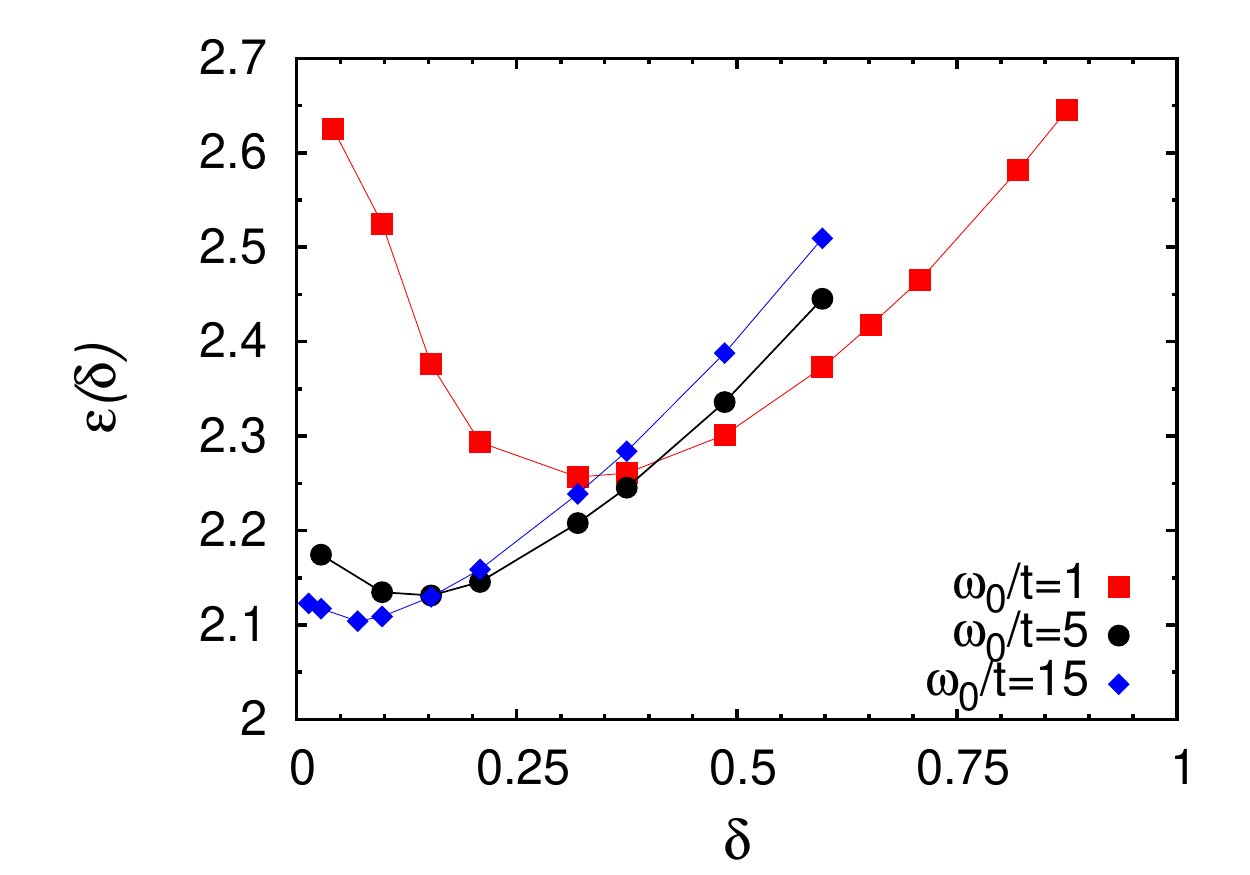}
\caption{\label{fig:phasesep3}
Energy per hole of Eq.~(\ref{eq:emery}) for $U=0$ and $\lambda/t=2$ for $\omega_0/t=1$ (red squares), $5$ (black circles), and $15$ (blue 
diamonds). Calculations are performed on the $12 \times 12$ cluster with TABC. Errorbars are smaller than the size of the symbols.}
\end{figure}

Finally, we compare the energy per hole for $U=0$ and $\lambda/t=2$ for different values of $\omega_0/t$, see Fig.~\ref{fig:phasesep3}. 
In all these cases, the ground state at half filling has CDW order (see Fig.~\ref{fig:phased}) and, therefore, phase separation is expected 
to appear away from half filling. However, in the antiadiabatic limit $\omega_0/t \to \infty$, there is no phase separation, since the 
Holstein model maps to the negative-$U$ Hubbard model, which has a uniform ground state away from half filling. In fact, we find that phase 
separation reduces when increasing $\omega_0/t$, i.e., the position of the minimum in the energy per hole shifts toward $\delta=0$, indicating 
that our variational approach correctly reproduces the expected physical behavior.

%%%%%%%%%%%%%%%%%%%%%%%%%%%%%%%%%%%%%%%%%%%%%

%%%%%%%%%%%%%%%%%%%%%%%%%%%%%%%%%%%%%%%%%%%%%

\section{Conclusions}\label{sec:conclusions}

In conclusion, we have performed accurate VMC calculations to extract thermodynamic properties of the Hubbard-Holstein model, where 
finite-size effects have been strongly reduced by implementing an average over twisted boundary conditions. At half filling, our results 
confirm the existence of a gapless phase between the CDW and the antiferromagnetic insulators, as recently obtained by different VMC 
calculations~\cite{Ohgoe2017}. Moreover, within our approach, which is based upon a transparent parametrization of the variational wave 
functions, we are able to observe the presence of superconducting correlations in the intermediate phase. When the phonon energy becomes 
large, pairing correlations strengthen and the superconducting region broadens to the detriment of CDW order. The emergence of 
superconductivity in the half-filled Hubbard-Holstein model is an example on how two competing tendencies (i.e., antiferromagnetism, 
favored by electron-electron interaction, and CDW order, favored by electron-phonon coupling) may lead to a third stable phase. In addition, 
we studied the effect of hole doping for both regimes where the half-filled ground state has either CDW or superconducting order. In the 
former case, a substantial phase separation is present at small dopings, resembling the case of a doped repulsive-$U$ 
Hubbard~\cite{Macridin2006b,Aichhorn2007,Chang2008,Misawa2014,Sorella2015,Tocchio2016,Simkovic2017}. In the latter case, instead, the 
ground state remains uniform with superconducting order. However, superconductivity is found to monotonically decrease upon doping. 
We remark that, away from half filling, TABCs are fundamental to reduce finite-size effects.

From general grounds, within the Hubbard-Holstein model, superconductivity is rather fragile against electron-electron repulsion and also
against electron doping. Indeed, since phonons are coupled to the local electronic density in the Hubbard-Holstein model, there is a direct 
competition between the formation of superconducting pairs and the local Coulomb repulsion $U$. In addition, superconducting pairing
is maximum at half filling and strongly reduces in the presence of hole doping. In this respect, a different scenario is expected within the 
Su-Schrieffer-Heeger model~\cite{Su1979}, where lattice displacements are coupled to the hopping term: here, no superconductivity is 
expected at half filling, since a Peierls insulator should take place for any electron-phonon coupling at $U=0$ (similarly to what happens 
in one dimension~\cite{Fradkin1983,Baeriswyl1985,Weber2015}). On the contrary, superconductivity is expected to emerge upon doping, being 
also more robust against Coulomb repulsion than in the Hubbard-Holstein model. Therefore, the Su-Schrieffer-Heeger model would provide 
a different mechanism for electron pairing, more pertinent for cuprate and iron-pnictide superconductors. Further variational investigations 
in this direction could benefit from the use of backflow terms, as introduced to improve the quality of the wave functions 
in the Hubbard model~\cite{Tocchio2008,Tocchio2011}.

\acknowledgments

We thank M. Fabrizio for useful discussions as well as T. Ohgoe and M. Imada for sharing their VMC numerical data.

\end{document}